\documentclass[%
 reprint,
 amsmath,amssymb,
 aps,
prb,
]{revtex4-2}

\usepackage{soul}
\usepackage{enumitem}
\usepackage{amsfonts}
\usepackage{bbm}

\usepackage{graphicx}
\usepackage{dcolumn}
\usepackage{bm}
\usepackage{xcolor}

\def\br{{\bf r}}

 \usepackage{braket}

\begin{document}

\preprint{APS/123-QED}

\title{Self-consistency in $GW\Gamma$ formalism leading to quasiparticle-quasiparticle couplings}
\author{Carlos Mejuto-Zaera}
\email{carlos\_mejutozaera@ucsb.edu, current: cmejutoz@sissa.it}
\affiliation{University of California, Santa Barbara}
\author{Vojt\v{e}ch Vl\v{c}ek}
\email{vlcek@ucsb.edu}
\affiliation{University of California, Santa Barbara}

\date{\today}
\begin{abstract}
    Within many-body perturbation theory, Hedin's formalism offers a systematic way to iteratively compute the self-energy $\Sigma$ of any dynamically correlated interacting system, provided one can evaluate the interaction vertex $\Gamma$ exactly. This is however impossible in general, for it involves the functional derivative of $\Sigma$ with respect to the Green's function. Here, we analyze the structure of this derivative, splitting it into four contributions and outlining the type of quasiparticle interactions that each of them generate. Moreover we show how, in the implementation of self-consistency, the action of these contributions can be classified into two: a quantitative renormalization of previously included interaction terms, and the inclusion of qualitatively novel interaction terms through successive functional derivatives of $\Gamma$ itself, neglected until now. Implementing this latter type of self-consistency can extend the validity of Hedin's equations towards the high interaction limit, as we show in the example of the Hubbard dimer. Our analysis also provides a unifying perspective on the perturbation theory landscape, showing how the T-matrix approach is completely contained in Hedin's formalism.
\end{abstract}
\maketitle

Many-body perturbation theory (MBPT) has become widely applied to obtain the description of quasiparticle (QP) energies, in particular the electron and hole states, and their couplings with fields~\cite{Onida2002}, with promising perspective for non-equilibrium phenomena~\cite{Perfetto2022}. Fundamentally, the perturbation theory framework relates the fully interacting many-body system to a suitably chosen auxiliary (often fictitious) reference one. The latter is selected so that the property of interest (e.g., the spectral function $A(\omega)$ representing the QP energies) is easy to evaluate and close to that of the physical interacting system. The ``closeness'' is typically defined as a perturbation series in terms of a parameter, which is small enough to guarantee the convergence and physical accuracy. A successful treatment hinges on the ability to generate the complete perturbative series and practically evaluate it~\cite{Abrikosov1975, martin_reining_ceperley_2016}. 

The MBPT framework is formulated around the one-body Green's function $G$, which is directly related to physical observables, e.g. $A(\omega) \propto {\rm Im} G(\omega)$~\cite{Hufner1984,Sander1987,Chewter1987,Puschnig2009,Vozzi2011,Wu2011,Luftner2014,Zhou2020}. The reference $G_0$ is usually computed from a system subject to mean-field interactions. The perturbative correction is then defined via the Dyson equation $\Sigma = G_0^{-1} - G^{-1}$, with the self-energy (SE),  $\Sigma$, being the central quantity leading to dynamical renormalization. Conceptually, $\Sigma$ corresponds to an effective scattering potential for the reference QPs, which itself functionally depends on $G$~\cite{martin_reining_ceperley_2016,Onida2002}. Further, it recovers the full many-body $G$ in the infinite resummation limit of the Dyson expansion. In practice, multiple approximations to the form of $\Sigma$ have been explored and applied in various contexts~\cite{delSole1994,Shishkin2007, Hellgren2018, Springer1998,Bruneval2013,Blase2011,Vlcek2017,ZhangD2017,Vila2020,Tzavala2020,Vlcek2018-PRM,Lischner2015}. Typically, these complementary Ans\"atze are based on a choice of the leading scattering mechanism combined with a particular closure of the perturbative expansion. These various methods are often considered to originate from distinct philosophies for generating $\Sigma$ from the two-particle interactions, how the expansion closure is implemented, and whether self-consistency is necessary.

Among these methods, we revisit here Hedin's approach~\cite{Hedin1965,Aryasetiawan1998}, which has been traditionally tied to the context of weak interactions dominated by classical electrodynamic screening and cases when explicit two-particle interactions are merely second (and higher) order effects. We challenge this notion and show that it intrinsically contains the complete framework necessary to generate any type of $\Sigma$ via a \textit{functional} self-consistency. This has been recognized but explored in only a very limited way  so far~\cite{Schindlmayr1998,Grueneis2014,Kutepov2017,Vlcek2019,Wang2021}, and here we show exactly how it stands apart from the numerical self-consistent solutions for a fixed form of $\Sigma$. 

Hedin's equations provide a systematic approach to building $\Sigma$ for a chosen $G_0$~\cite{Hedin1965,Aryasetiawan1998,martin_reining_ceperley_2016}, relating $G$ to $G_0$ and the bare Coulomb potential $v$ through the self-energy $\Sigma$ and the so-called interaction vertex $\Gamma$. The latter term dresses the screened Coulomb potential $W$ with the necessary many-body interactions to form $\Sigma$ following

\begin{equation}\label{eq:SE}
    \Sigma(1,2) = iG(1,\bar 3)W(1,\bar 4)\Gamma(\bar 4,\bar 3,2).
\end{equation}

We adopt a short hand notation where space-time coordinate $1 \equiv (\br_1,t_1)$; all coordinates integrated over are indicated by bar. Hedin's equations are closed by interrelating all quantities, and this set of coupled equations should be solved self-consistently. In this work, we focus on the role of vertex function in this self-consistency. It is directly given in terms of $\Sigma$ as a resummation:
\begin{equation}\label{eq:gamma}
    \Gamma(1,2,3) = \delta(1,2)\delta(1,3) + \frac{\delta \Sigma (1,2)}{\delta G(\bar 4, \bar 5)} G(\bar 4,\bar 6) G(\bar 7,\bar 5) \Gamma (\bar 6,\bar 7,3).
\end{equation}
We denote the functional derivative in the RHS as the interaction kernel $\mathcal{K}(1,2,3,4)= \delta \Sigma(1,2) /\delta G(3,4)$ . 

The full self-consistent solution requires that the SE in the $n^{\rm th}$ iteration $\Sigma^{(n)}$ enters Eq.~\ref{eq:gamma} and generates a new form of the vertex $\Gamma^{(n+1)}$ via $\mathcal K^{(n)} [\Sigma^{(n)}]$. At each additional step, this produces a SE containing \textit{new} diagrams, making any exact evaluation difficult. We emphasize that this is distinct from a typical self-consistent treatment, in which the functional form of $\mathcal{K}$ is not updated and only numerical convergence is sought. The functional self-consistency between $\mathcal{K}$ and $\Sigma$ accounts for a resummation in types of many-body interactions. 

Practical implementations resort to imposing closure to Hedin's formalism. We identify two typical approaches, which we will explore here: (i) truncation of the $\Sigma$ expansion at finite order in interactions or (ii) restricting the appearing diagram topologies that are summed over to all orders. Further, the numerical self-consistency is also typically avoided to lower the computational cost.

\begin{figure}
    \centering
    \includegraphics[width=0.45\textwidth]{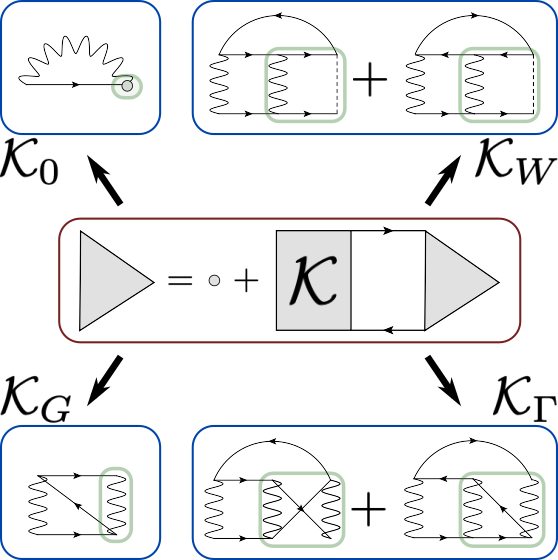}
    \caption{Components of the interaction kernel $\mathcal{K}$ in the interaction vertex $\Gamma$ (red box), with diagrams representing the leading order contributions to the self-energy (blue boxes).}
    \label{Fig:Gamma}
\end{figure}

First, we illustrate contributions of the kernel $\mathcal{K}=\delta (GW
\Gamma)/ \delta G$ to both $\Gamma$ and $\Sigma$ and impose the truncation: we select the lowest order terms in bare and screened Coulomb interactions ($v$ and $W$) take only the first non-trivial step in the resummation of Eq.~\ref{eq:gamma}. The corresponding Feynman diagrams for the SE are in Fig.~\ref{Fig:Gamma}, in which the corresponding $\mathcal K$ parts are indicated by green rectangles. We distinguish four types of kernels based on how they are generated, and interpret them in terms of scattering processes.

\begin{enumerate}[wide, labelwidth=!, labelindent=0pt]

    \item $\mathcal{K}_0=0$: This is the trivial term in Eq~\ref{eq:gamma} corresponding to $\Gamma(1,2,3) = \delta(1,2)\delta(1,3) $ and leading to the zeroth order approximation for the SE:
    \begin{equation}
    \Sigma^{(0)}(1,2) = iG(1,2)W(1,2)    
    \end{equation}
    It accounts exclusively for a classical, 1-body screened Coulomb interaction,\footnote{containing a resummation over polarization diagrams} i.e., the screened exchange diagram in Fig.~\ref{Fig:Gamma}. While greatly improving main QP mean-field features over mean field (e.g., DFT) results~\cite{Hybertsen1985,Hybertsen1986,Godby1986}, it suffers form self-polarization error in $W$~\cite{Nelson2007, Romaniello2009, Aryasetiawan2012}.
    
    \item $\mathcal{K}_G = i(\frac{\delta G }{\delta G})W\Gamma$, arising from the variation of $G$. To generate a nontrivial vertex, it is sufficient to use the zeroth order approximation to $\Gamma$. Hence 
    \begin{equation}
    \mathcal{K}^{(0)}_G (1,2,3,4)= iW(1,2)\delta(1,3)\delta(2,4).\label{eq:Kernel0G}
    \end{equation}
    This introduces ``ladder'' interactions in $\Gamma^{(1)}$ which enter the recurrence in Eq.~\ref{eq:gamma}.
    \begin{equation}
        \Gamma^{(1)}_G (1,2,3)= i G(1,3)W(1,2)G(2,3)
    \end{equation}
    where we induce a subscript to indicate from which kernel the vertex is derived. To the lowest order in interaction lines ($W$), the vertex-corrected SE becomes:
    \begin{align}\label{eq:SE1_KG}
    \Sigma^{(1)}(1,2) =&  i G(1,\bar 3) W(1,\bar 4) \bigg[\delta(\bar 3,2)\delta (\bar 4, 2) \nonumber \\ &+  i G(\bar 3, \bar 4)W (\bar 3, 2)G(\bar 4, 2)\bigg].
    \end{align}
     The higher order expansions are illustrated in the SI. Eq.~\ref{eq:SE1_KG} corresponds to the two left-most SE diagrams in Fig.~\ref{eq:gamma} based on $\mathcal K_0$ and $\mathcal K_G$ with screened Coulomb interactions. In practice, $\mathcal K_G$ introduces additional exchange coupling between the virtual particle-hole (ph) pairs and accounts for excitonic effects. Further, it partly corrects the self-polarization error~\cite{Grueneis2014,Maggio2017,Vlcek2019,Mejuto2021}.
    
    \item $\mathcal{K}_W =iG ( \frac{\delta W }{\delta G})\Gamma$, arising from the variation of the screened Coulomb potential.  Here again, we generate the lowest non-trivial vertex already from $\Sigma^{(0)}$.  To the lowest order, $W$ is given by the random phase approximation:
    $W(1,2) = v(1,2) -i v(1,\bar 3)  G(\bar 3,\bar 4)G(\bar 4,\bar 3) W(\bar 4,2)$ and in the first step of Eq.~\ref{eq:gamma} resummation, it yields:
    \begin{align}\label{eq:Gamma1W}
        \Gamma^{(1)}_W(1,2,3) &= v(1,\bar 4)G(1,2)W(2,\bar 5) \nonumber \\ \times \bigg[& G(\bar 4,\bar 5) G(3,\bar 4)G(\bar 5, 3) +G(\bar 5,\bar 4) G(3,\bar 5)G(\bar 4, 3) \bigg]
    \end{align}
    The vertex represents the second-order correction in terms of interactions ($v$ and $W$). Note that $\Gamma^{(1)}_W(1,2,3)$ contains explicit particle-particle (pp) and particle-hole (ph) interactions at equal footing~\cite{Schindlmayr1998}.
    
    As a result, the lowest order SE in Fig.~\ref{Fig:Gamma} contains 2-body terms, which cannot be reduced to a single propagator interacting with itself through a screened potential $W$. The leading order terms in $\Sigma$ from $\Gamma_W$ are identical to the direct terms found in the \textit{screened} T-matrix expansion~\cite{Springer1998,Romaniello2012,Nabok2021}.  The higher order screened T-matrix is outside of the imposed truncation and results from the second type of closure (resummation restricted to a selected type of diagrams) discussed below.
    Further, note that it is common to take $\delta W/\delta G \to 0$ as the change of screening is expected to be small, typically in large scale systems or systems with low screening where $W \sim v$. One is thus tempted to take the limit of $ W\to v$ prior to evaluating the vertex. However, to keep the explicit 2-body couplings in the SE expression, such a limit has to be taken only \textit{after} $\Gamma$ has been derived in each step.

    \item $\mathcal{K}_\Gamma = iGW (\frac{\delta \Gamma }{\delta G})$ comes from the variation of the vertex in the kernel, which has been, to the best of our knowledge, neglected so far. A non-vanishing  $\mathcal{K}_\Gamma$ requires a nontrivial vertex as an input. Clearly, this step can generate an infinite number of diagrams given that $\Gamma$ is subject to the resummation in Eq.~\ref{eq:gamma}. By including only the leading term, we get:
    \begin{align}
    \frac{\delta}{\delta G(4,5)} \Gamma(1,2,3) &=  \frac{\delta}{\delta G(4,5)}\bigg[\mathcal{K}(1,2,\bar 6,\bar 7) G(\bar 6,3) G(3,\bar 7)\bigg]\nonumber \\
    &= \frac{\delta\mathcal{K}(1,2,\bar 6,\bar 7)}{\delta G(4,5)} G(\bar 6,3) G(3,\bar 7)\nonumber\\
    &+\mathcal{K}(1,2,4,\bar 7) G(3,\bar 7) \delta(3,5) \nonumber \\
    &+\mathcal{K}(1,2,\bar 6,5) G(\bar 6,3)\delta(3,4). 
    \end{align}
    Since we are interested only in the lowest order expression in interactions (i.e., with the lowest number of $v$ or $W$ terms), we take $\mathcal K \approx \mathcal{K}^{(0)}_G = iW$ and further $\delta  \mathcal{K}^{(0)}_G /\delta G \approx 0$. As a result,  
    \begin{align}
        \Gamma^{(2)}_{\Gamma}(1,2,3) &=  - W(1,\bar 4)G(1,\bar 5)W(2,\bar 5) \nonumber \\ \times \bigg[& G(\bar 5,3) G(3,\bar 4)G(\bar 4, 2) +G(\bar 5,\bar 4) G(\bar 4, 3) G(3,2)\bigg]
        \label{eq:Gamma_Gamma}
    \end{align}
    In practice, this expression is analogous to Eq.~\ref{eq:Gamma1W}  upon exchange of space-time coordinates. A significant difference is the absence of $v$ terms, i.e., the vertex contains only screened interactions. In the SI, we show how this difference is resolved if one resums the formal Dyson-like equation that defines $\frac{\delta W}{\delta G}$. Note that the $\Gamma^{(2)}_\Gamma$ represents the same order of perturbation expansion, as $\Gamma_W^{(1)}$; it is important that the superscript does not denote the order of the expansion, but enumerates the iteration in the self-consistent cycle. Clearly, the number iterative steps taken in Eq.~\ref{eq:gamma} is not equivalent to the order of the perturbation expansion.
    
    Finally, the resulting SE is in Fig.~\ref{Fig:Gamma} and corresponds to the exchange form of the the T-matrix~\footnote{This is in complete analogy to how the Fock diagram can be constructed from the Hartree term; hence we refer to this topological analogy as exchange-equivalent diagram}. Here, $\Gamma^{(2)}_\Gamma$ corrects the 2-body pp and ph interactions in $\Gamma_W$ by accounting for the Fermionic nature of the QPs. 
\end{enumerate}

In the above steps, we derived the leading order terms where we selectively limited the expansion of $\Gamma$ up to the second order in $v$ and $W$. This constitutes a particular form of a closure, i.e., finite order truncation. Using this closure, we generated a self-energy expression containing (lowest order) direct and exchange T-matrix terms for both pp and ph channels. In contrast, such diagrams were previously generated by imposing an Ansatz for the two-body interactions and typically only one of the channels, pp or ph, was applied~\cite{Springer1998,Romaniello2012,Nabok2021}. Our derivation shows that both need to be added simultaneously and at equal footing.  
All these diagrams are shown in Fig.~\ref{Fig:Gamma}.

\begin{figure}[h!]
    \centering
    \includegraphics[width=0.49\textwidth]{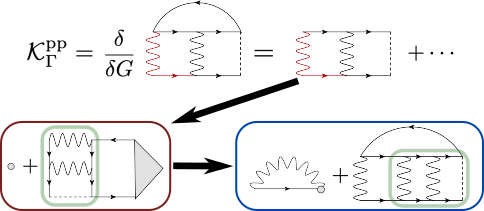}
    \caption{Graphical sketch for the generation of topologically new diagrams through the functional self-consistency involving the $\Gamma$ component of the interaction kernel $\mathcal{K}$. On the first line, $\Gamma$ portion of the lowest order pp T-matrix-equivalent SE is in black, while the $GW$ component is shown in red. In particular, generation of the second order direct pp T-matrix diagram from the first order term. The red box encloses the interaction vertex $\Gamma$, the blue box the resulting self-energy $\Sigma$ representing the next lowest order T-matrix SE.}
    \label{fig:Tmtx_higher}
\end{figure}

Continuing the derivation towards full self-consistency (updating the kernel), $\mathcal K$ in Eq.~\ref{eq:gamma} will generate a large number of additional diagrams. There is a marked difference in how each non-trivial vertex does this, hinging on whether and how the functional derivative of $\Gamma$ (the defining component of $\mathcal{K}_\Gamma$) enters their expression. Indeed, $\mathcal{K}_G = W\Gamma$ at each iteration. 
In turn, $\mathcal{K}_W$ is composed of three terms, two of which do not contain $\delta \Gamma/\delta G$. Hence, $\mathcal{K}_G$ and two out of the three terms in $\mathcal{K}_W$ contribute with diagrams of a particularly restricted structure, essentially further renormalizing the same interaction channels in each iteration (c.f. with the ``starfish'' algorithm in Ref.~\cite{Tandetzky2015}). In strong contrast, the terms involving $\delta \Gamma/\delta G$, i.e. one of the terms in $\mathcal{K}_W$ as well as the entire $\mathcal{K}_\Gamma$, can introduce more complex diagram topologies in each iteration. As we illustrate below, these functional derivatives are critical for expanding the series of explicit QP-QP interactions.

For instance, evaluating the $\mathcal{K}_\Gamma$ kernel as $GW({\delta\Gamma_W^{(1)}}/{\delta G})$ leads to the next order T-matrix, shown in Fig.~\ref{fig:Tmtx_higher}, along with three additional diagrams that are not shown. Clearly, if the full (functional) self-consistency is sought, the vertex acts as a ``generator'' of new diagram topologies and $\delta \Gamma/\delta G$ is constantly expanding with each additional step in the series. Conceptually, this self-consistency potentially extends the convergence radius of the PT expansion, by introducing additional types of diagrams with each iteration. The way the new diagrams arise through $\mathcal{K}_W$ and $\mathcal{K}_\Gamma$ is subtly different, however, $\mathcal{K}_W$ essentially modifying the effective potential through which a given QP interacts with the background particles and holes, while $\mathcal{K}_\Gamma$ can also affect the propagation itself of said QP. This is likely the reason behind previous numerical observations regarding the absence of improvement over $GW$ in approximations including vertex corrections exclusively in the polarization function~\cite{Lewis2019}.

It is at this point that the second type of closure becomes apparent: the expansion is formally continued ad infinitum  but restricted to a subset of diagrams at each iteration. For example, if we restrict the self-energy entering $\mathcal K$ to contain only $\Gamma_W$-type vertex terms, it consistently generates a series of diagrams among which we can always identify one that produces the next leading order pp and ph T-matrix, as illustrated in Fig.~\ref{fig:Tmtx_higher} for the case of pp T-matrix. This analysis demonstrates the critical role of $\frac{\delta \Gamma}{\delta G}$ terms, particularly within $\mathcal{K}_\Gamma$, as one cannot generate such a series if they are neglected.

Note that this repeated update of the functional form of $\Gamma$ is distinct from the resummation of Eq.~\ref{eq:gamma} which does not generate the T-matrix type of self-energy. For instance, in the context of the lowest order T-matrix ladder diagram (upper right corner in Fig.~\ref{Fig:Gamma}), Eq.~\ref{eq:gamma} yields a cascade of \textit{single} pp or ph scatterings. In contrast, the above described self-consistency generates a series of repeated pp and ph scattering events that represent the (screened) T-matrix (cf. Fig.~\ref{fig:Tmtx_higher}).

The previous derivation reaches two goals. First, we have shown what types of interactions the leading order terms of $\mathcal{K}$ introduce in Hedin's construction of the self-energy. Moreover we show that this formalism naturally generates self-energy containing T-matrix types of diagrams, without needing to invoke the Bethe-Salpeter equation. This exemplifies the role of the recurrence in Eq.~\ref{eq:gamma}, and provides hence a unified framework from which to systematically derive and analyze previously developed vertex corrected MBPT approximations~\cite{Romaniello2012,Nabok2021}, which have targeted particular phenomenologies. For instance, our derivation indicates that \emph{both} pp and ph T-matrix terms have to be added together into a comprehensive $\Sigma$ when attempting to apply MBPT systematically. 
As shown above, practical implementation relies on an appropriate closure. 

Naturally, a questions arises about what approach is better in practice and how individual vertex terms change the prediction of QP energies. Further, the fact that vertex terms introduce QP-QP interactions suggests that the perturbative treatment may better capture systems with stronger interactions, but the practical limit of the perturbation expansion is unclear. Finally, regardless of the selected closure, one should perform a fully self-consistent calculation which should, in principle, become independent of the starting point. 

In the following, we address these questions with self-consistent calculations of a minimal model, the Hubbard dimer at half filling, which became de facto standard test case for MBPT~\cite{Romaniello2009,Romaniello2012,DiStefano2021}. While this numerical test is far from exhaustive, it already shows the main qualitative features and  demonstrates the role of various vertex terms and types of closures.

\begin{figure}
    \centering
    \includegraphics[width=0.5\textwidth]{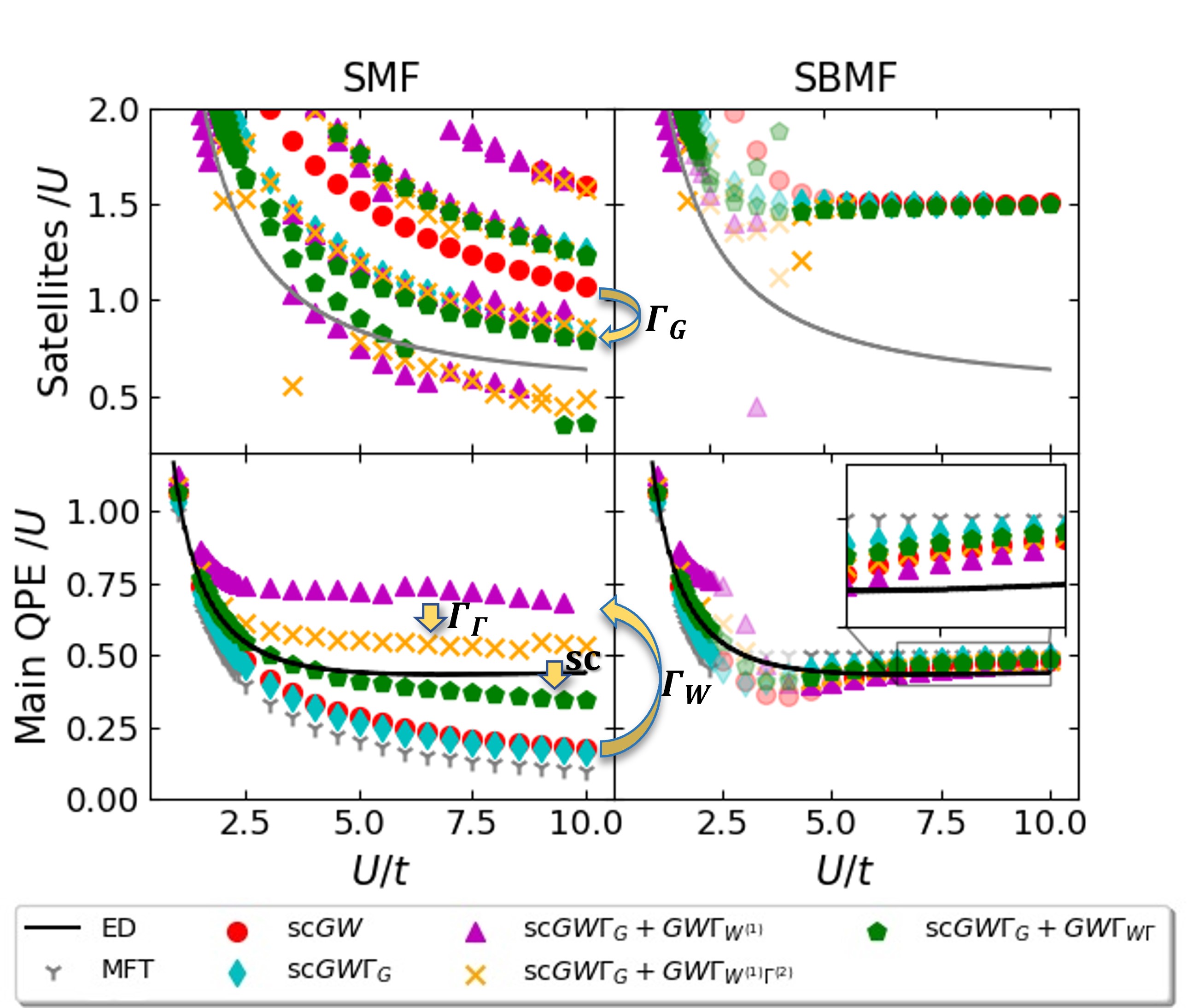}
    \caption{Quasi-particle energies (QPEs) from the spectral function $A(\omega)$ for the Hubbard dimer at half-filling for different $U/t$ values, with main QPEs in the lower panels, satellites in the upper panels. The QPEs are rescaled by $U$. Shown are ED results as solid lines (main QPE in black, satellite in grey), as well as mean-field and several MBPT approximations as solid markers. Transparent markers in the right panels correspond to calculations where a stable convergence was not possible. Results for a translational symmetric mean-field (SMF) are shown on the left panels, and a symmetry broken mean-field (SBMF) on the right panels. Arrows indicate the effect of the different vertex corrections, and the functional self-consistency (sc), see text for details.}
    \label{fig:QPE_comp}
\end{figure}

We perform MBPT calculations on the Hubbard dimer at half filling for different $U/t$ ratios in the span between 1 and 10, starting from a translational invariant (i.e., ``symmetric'') Hartree-Fock mean-field reference (SMF), as well as allowing for translational symmetry breaking (SBMF). As discussed in the SI, the changing point between the delocalized and atomic behaviours in the dimer occurs at $U/t = 2$, hence the $U/t \in \left[1,10\right]$ are sufficiently large to study both the weakly and strongly interacting regimes. We present the main results in Fig.~\ref{fig:QPE_comp}, which shows the QP energies rescaled by $U$ as a function of $U/t$ for different perturbative methods and mean-field references~\footnote{We identify the main QP peak as the one continuously evolving from the only spectral feature in the $U/t = 0$ limit upon increase of the Coulomb interaction.}. For clarity, main QP energies are shown in the lower panels, satellites in the upper panels. We compare to exact diagonalization (ED) results shown as solid lines (black for the main QP, grey for the satellite). The details of the mean-fields and perturbative calculations can be found in the SI.

We employ two reference MF states: SMF and SBMF. Ideally, in a fully self-consistent and resummed MBPT implementation, there should be no dependence on the starting point. However, since we introduce limited types of diagrams and partial self-consistency (see below), this factor cannot be neglected. As we show in detail in the SI, the MF solution for the Hubbard dimer at half-filling is unstable to symmetry breaking for $U/t>2$, corresponding to the atomic limit. The symmetry broken solution greatly improves the MF total energy, following almost the exact asymptotic behavior for $U/t\to\infty$ and reasonably reproducing the main QP $A(\omega)$ (grey stars in Fig.~\ref{fig:QPE_comp}). Since the atomic limit is perturbative around $t^2/U$, the question arises whether SBMF or SMF would be a better reference state for MBPT in the high $U/t$ limit. On the one hand, the energetic closeness of the SBMF to ED is encouraging, but it displays spurious anti-ferromagnetic order. The exact self-energy would eliminate it, yet only by including diagrams with spin-flip processes. This limits the usefulness of the SBMF starting point with our current MBPT implementations.

Indeed, as can bee seen comparing both panels in Fig.~\ref{fig:QPE_comp}, the main QPE (shown as solid black line for ED) is much more faithfully reproduced in the high $U$ limit by any MBPT approximation starting from SBMF instead of SMF, while in the intermediate $U$ regime we found trouble converging the different approximations, which we indicate with semi-transparent markers in the Figure. However, upon closer inspection (see inset in lower right panel), one notices that essentially none of these methods are significantly improving upon the mean-field reference. Satellite features do appear, but their energy is gravely overestimated and their asymptotic distance to the main peak is qualitatively wrong ($\sim U$ instead of constant). The reason behind the lack of improvement is likely the broken symmetry. All the interaction terms and diagrams we include conserve local spin, and thus none of our methods can recover the broken symmetry~\footnote{Which would likely be remedied in a relativistic framework.}. This suggests that the SBMF solution is the optimal symmetry broken description of the model at the high $U$ limit, and we expect a perturbative improvement within the same magnetic phase to be unlikely~\footnote{Alternatively, one could include a symmetry breaking term $h$ into the MBPT expansion, and take $h\to0$ at the end of the calculation. Such a driven solution could overcome this issue, in principle, but it requires careful treatment.}. Further, we  interpret the the absence of multiple satellites for the symmetry broken case
(Fig.~\ref{fig:QPE_comp}, right upper panel ) in our MBPT 
implementations as another sign that the MF solution is the optimal 
representation in the SB space. Essentially, none of the perturbative approximations are moving significantly far from the original mean-field solution in the symmetry broken case, resulting in a much simpler, albeit incorrect, satellite structure. Given this limitation, for the rest of the paper we will concentrate exclusively on the results starting from the SMF solution.

We first consider the lowest order approximations, performed to self-consistency (sc): sc$GW$ (red circles) and sc$GW\Gamma_G$ (cyan diamonds). sc$GW$ opens the main QP gap from the SMF solution, which is fix at $2t$ independently of $U$. This is the oposite tendency as observed in most \emph{ab inito} materials~\cite{Aryasetiawan1998,Onida2002,Golze2019}, and is a consequence of the Hartree and Fock interactions cancelling exactly in the Hubbard model. Furthermore, sc$GW$ introduces satellites that systematically overestimate the ED ones (grey solid line), but generally follow a similar behaviour with $U/t$. Upon the introduction of induced density matrix fluctuations and partial removal of the self-polarization error with the sc$GW\Gamma_G$ approximation, we see these satellite features significantly reduce their energy, coming much closer to the ED result. Meanwhile, the main QP energies are barely affected by the introduction of the $\mathcal{K}_G$ kernel. As we will show next, this trend is repeated at all levels of perturbation theory: the $\mathcal{K}_G$ kernel, and only this kernel, affects the satellites, leaving the main QP features intact. While for systems with more general interaction terms one would expect some effect on the main QP energies, these findings explain the numerical observations made previously~\cite{Grueneis2014,Maggio2017,Vlcek2019,Mejuto2021}.

Next, we include the $\mathcal{K}_W$ kernel to leading order (c.f. Eq.~\eqref{eq:Gamma1W}), which we apply in the $W\to v$ limit as a one-shot correction to the sc$GW\Gamma_G$ self-energy, shown as magenta triangles in Fig.~\ref{fig:QPE_comp}. We observe that the satellites remain mostly unchanged from sc$GW\Gamma_G$, up to the appearance of additional spurious satellites at high energy. In strong contrast, the main QP energies experience a huge shift to higher energies, particularly in the large $U/t$ limit. This is not surprising, since it is in this regime where the pp and ph interaction diagrams from $\mathcal{K}_W$ become dominant. While the main QP energies are clearly overestimated, there is an encouraging change in the asymptotic behaviour with $U/t$, resembling much stronger the correct flat ($\sim U$) one.

The systematic overestimation of the QP energies gets significantly corrected by recovering the exchange diagrams in $\mathcal{K}_\Gamma$. We evaluate these also to leading order (c.f. Eq.~\eqref{eq:Gamma_Gamma}) in the $W\to v$ limit, and add them as one-shot correction to the self-energy (yellow crosses in Fig.~\ref{fig:QPE_comp}). This implementation keeps the promising asymptotic behaviour (and satellite peaks), while pushing the main QP energies much closer to the ED results. It is in this sense that we extend the qualitative validity of MBPT towards the high interaction limit by including diagrams through $\mathcal{K}_\Gamma$. Further, this indicates that we are adding the most important missing interactions in the SMF model, namely the explicit two-body (exchange corrected) Coulomb interactions, in \emph{both} pp and ph channels. This compares favorably with previous studies within the T-matrix approach, which could not observe this improvement in the large $U/t$ limit since they added either the pp or the ph channel, but not both together~\cite{Romaniello2012}.

Finally, we want to illustrate the effect of the self-consistency in $\mathcal{K}_\Gamma$, namely the introduction of a full resummation of a subset of diagrams which arise naturally through $\frac{\delta \Gamma}{\delta G}$ in successive iterations of Hedin's equations. Here, we resum all T-matrix type diagrams, which we add as one-shot correction to the sc$GW\Gamma_G$ self-energy (green pentagons in Fig.~\ref{fig:QPE_comp}), accounting thus for the leading order terms in all kernels $\mathcal{K}$, as well as both types of self-consistencies (numerical and functional). The satellite features improve slightly towards the ED limit, but the most important change is undoubtedly the effect on the main QP energies. These become essentially exact up to $U/t = 4$, showing an extension of the qualitative \emph{and} quantitative validity of MBPT towards the high interaction limit, going through the formally ``strongly correlated'' point at $U/t = 2$.

Still, the persistent satellite overestimation and  appearance of spurious satellites suggest that renormalization of the new scattering channels to the correct quantitative energy scale is still missing. Moreover, as can be seen at the main QP energies for $U/t > 5$, further diagrams are likely needed to recover the exact results. This may require resummation within the diagram classes already introduced, such as by including the T-matrix-like $\Sigma$ into the self-consistency procedure, including the $\mathcal{K}_W$ and $\mathcal{K}_\Gamma$ contributions to $W$, or introducing higher order functional derivative terms in $\mathcal{K}$. We attempted including both the leading order and resummed versions of $\Sigma[\Gamma_W,\Gamma_\Gamma]$ into the self-consistency procedure (see SI), but the resulting MBPT method does not converge to a stable Green's function for $U/t > 5$, in contrast to the $U/t <5$ regime.

Nonetheless, the numerical results support our theoretical conclusions: The higher order functional derivative terms in $\mathcal{K}_\Gamma$ introduce new classes of diagrams, which are necessary to extend the qualitative applicability of MBPT towards the high interaction limit. In other words, $\mathcal{K}_\Gamma$ affects the convergence radius of the perturbative expansion. Meanwhile, resummations within a diagram class, be it within the solution of a Dyson-like equation or in the self-consistency of $\mathcal{K}_0$, $\mathcal{K}_G$ and $\mathcal{K}_W$, fixes the quantitative agreement and hence is important for the reliability of the converged solution. 

For completeness, we conclude with a note of caution: As shown above, adding the $\mathcal{K}_{W,\Gamma}$ terms can extend the validity of Hedin-based MBPT from the weakly to the highly interacting limit. Still, if the system studied presents strong correlation between these two limits, e.g. at a phase transition, the Hedin expansion is not justified. Higher order perturbative corrections can only constrain the phase space region around strongly correlated points where MBPT is unreliable. Nevertheless, as we have shown in the Hubbard dimer, this can account for a significant portion of dynamical correlation in a numerically stable fashion.

In this study, we analyzed the structure of the interaction vertex $\Gamma$ in regards to its role as generator of quasiparticle interactions, and thus bridge between the mean-field and fully interacting Green's functions in Hedin's formalism for many-body perturbation theory. We identified the diagram types generated by each component of the interaction kernel $\mathcal{K} = \frac{\delta \Sigma}{\delta G}$, highlighting the effect of each one on the different features the experimentally accessible spectral function. Further, we outlined the two distinct types of self-consistency encountered in the formalism: That carried by the first elements of the kernel $\mathcal{K}_{0,G,W}$, which accounts for interaction renormalization and quantitative improvement of the description, and that governed by $\mathcal{K}_\Gamma \propto \frac{\delta \Gamma}{\delta G}$, neglected until now, which is responsible for generating new diagrams in each iteration, thus extending the validity of the perturbative approach through the phase diagram. Finally, this analysis allowed us to present a unified perspective on perturbative Ans\"atze, in particular showing how the T-matrix approach is generated within Hedin's formalism, and realizing the importance of including both particle-particle and particle-hole channels in the calculation. We hope this novel view on MBPT will help the development of scalable implementations towards the highly interacting limit, and look forward to the promising stochastic implementation of the $\mathcal{K}_{W,\Gamma}$ terms to large realistic systems~\cite{Vlcek2019}.

\section*{Acknowledgements}

Insightful discussions and comments from Arno F\"orster, Gabriel Kotliar, E. K. U. Gross, Jan Skolimowski and Michele Fabrizio are gratefully acknowledged. The theoretical analysis of vertex corrections was  supported by the NSF CAREER award (DMR-1945098).  The implementation, the detailed numerical analysis, and the preparation of the publicly available code is based upon work supported by the U.S. Department of Energy, Office of Science, Office of Advanced Scientific Computing Research, Scientific Discovery through Advanced Computing (SciDAC) program under Award Number DE-SC0022198. This research used resources of the National Energy Research
Scientific Computing Center, a DOE Office of Science User Facility
supported by the Office of Science of the U.S. Department of Energy
under Contract No. DE-AC02-05CH11231 using NERSC award
BES-ERCAP0020089.
\bibliography{CMZ_refs}

\end{document}